\documentclass{appolb}
\usepackage{epsfig}

\def\be{\begin{equation}}
\def\ee{\end{equation}}
\def\lsim{\raise0.3ex\hbox{$<$\kern-0.75em\raise-1.1ex\hbox{$\sim$}}}
\def\gsim{\raise0.3ex\hbox{$>$\kern-0.75em\raise-1.1ex\hbox{$\sim$}}}

\begin{document}
\pagestyle{plain}
\newcount\eLiNe\eLiNe=\inputlineno\advance\eLiNe by -1
\title{VECTOR MESON DOMINANCE
\thanks{Send any remarks to {\tt Dieter.Schildknecht@physik.uni-bielefeld.de}}%
}
\author{Dieter SCHILDKNECHT
\address{Fakult{\"a}t f{\"u}r Physik, Universit{\"a}t Bielefeld,
Universit{\"a}tsstrasse 25, 33615 Bielefeld, Germany}}
\maketitle

\begin{abstract}
Historically vector-meson physics arose along two different paths to be
reviewed in Sections 1 and 2. In Section 3, the phenomenological consequences
will be discussed with an emphasis on those aspects of the subject matter
relevant in present-day discussions on deep inelastic scattering in the
diffraction region of low values of the Bjorken variable.
\end{abstract}

\section{The gauge principle applied to properties of hadrons}
In the 1960ies, among particle theorists, there reigned the fairly
wide-spread opinion that the theory of strongly interacting particles
was to be formulated  in terms of the unitarity and analyticity properties
of the S-matrix by themselves, rather than relying on local quantum field
theory. All hadronic states being considered as equally elementary and at
the same footing, their dynamics was conjectured to be determined by the
``bootstrap conditions'' intensively put forward by G.F. Chew \cite{Chew}.

In his 1960 paper entitled ``Theory of strong interactions'', published in
Annals of Physics \cite{Sakurai}, J.J. Sakurai advocated an entirely 
different point of view.
Starting from the success of the principles of quantum field theory in 
quantum electrodynamics (QED), he emphasized that one should expect these
general principles to also hold for the physics of strong interactions.
In particular, the notion of conserved currents, the gauge principle
and the universality of couplings should be applied to strong-interaction
physics. Taking advantage  of the 1954 Yang and Mills generalization 
\cite{Yang} of 
U(1) gauge invariance from QED to local SU(2) gauge transformations, Sakurai
predicted the existence of vector mesons coupled to the hadronic isospin and
hypercharge currents. The predicted vector mesons were indeed experimentally
established in the years between 1961 and 1963, compare Table 1.
\begin{table}
\begin{center}
\begin{tabular}{|c|c|c|l|}\hline
I & Y & $J^P$ & \\ \hline
1 & 0 &     & $\rho$ (770) \\
0 & 0 & $1^-$ & $\omega$ (780), $\phi$ (1020) \\
1/2 & $\pm$ 1 & & $K^*$ (890) \\ \hline
\end{tabular}
\caption{The vector mesons}
\end{center}
\end{table}
Vector mesons played an important role in the generalization of 
$SU(2)_{isospin}$ to $SU(3)_{flavor}$. In ``The eightfold way: A theory of
strong interactions'' M. Gell-Mann said \cite{Gell}:

{\it ``The most attractive feature of the scheme is that it permits the
description of eight vector mesons by a unified theory of the Yang Mills
type (with a mass term). Like Sakurai, we have a triplet of vector mesons
coupled to the isospin current, ...''.}

The non invariance of the mass terms of the vector mesons was ignored for
the time being, local gauge transformations being considered as a means to
generate interactions with universal couplings among the nucleons and the 
vector mesons themselves.
For the $\rho$ meson triplet, for example \cite{Sakurai},
\be
f_{\rho} \equiv f_{\rho NN} = f_{\rho \pi \pi} = f_{\rho \rho \rho}.
\ee

In connection with the concept of mass, it may be appropriate to remind
ourselves of the situation at present. The mass problem, shifted to the 
masses of the leptons and quarks, even to-day still awaits the discovery
of the Higgs particle to be considered as being (partially?) solved.

In 1964, in ``A schematic model of baryons and mesons'' the three-dimensional
representation of (flavour) SU(3) became ``physical'' by introducing 
``quarks'' \cite{Gell-Mann}: 

{\it ``It is fun to speculate about the way the quarks would behave, if they
were physical particles of finite mass ...''.} (M. Gell-Mann).

Subsequently the quarks were endowed with an additional degree of freedom
\cite{Greenberg}
beyond electromagnetic charge. They became ``colored'', and the application of
the local gauge principle to the color degree of freedom of the quarks in
1972 led to $SU(3)_{Color}$ and Quantum Chromodynamics (QCD) \cite{Fritzsch}.

The vector mesons, now recognized as $(q \bar q)^{J=1}$ bound states, 
nevertheless find their
place as ``Dynamical gauge bosons of hidden local symmetry'' (Bando et al. 
1989 \cite{Bando}) in the framework of a low-energy effective Lagrangian
of massless two-flavored QCD, thus arriving at Sakurai's massive Yang-Mills
Lagrangian from a novel point of view.

\section{Electromagnetic interactions of hadrons.}
In the 1950ies the first measurements of the electromagnetic form factors
of the nucleons in electron-scattering experiments were performed. The 
interpretation of the form-factor measurements was the second path that
led to the existence of vector mesons.

Based on a picture of the nucleon as a core surrounded by a pion cloud, 
the form-factor measurements were interpreted as empirical evidence for an 
isoscalar vector meson, $\omega \to 3 \pi$, by Nambu
\cite{Nambu}  in 1957, and for an
isovector meson, $\rho^0 \to 2 \pi$, by Frazer and Fulco \cite{Frazer} in 1959.

Subsequently, this interpretation of the nucleon form factors was 
generalized to hold for the totality of all photon-hadron interactions,
formulated in terms of an operator identity \cite{GellIII,Kroll,N1}, 
known as current-field
identity (CFI). The electromagnetic current, the source of the Maxwell
field, $j^{elm}_\mu = J^{(3)}_\mu + {1 \over 2} J^{(Y)}_\mu$, was identified
with a linear combination of isovector and isoscalar vector-meson fields.
For e.g. the isovector part (with $m_\rho$ denoting the $\rho^0$-meson
mass, $\rho_\mu(x)$ denoting the $\rho^0$-meson field and $f_\rho
\equiv 2 \gamma_\rho$ the coupling), the CFI reads
\be
j^{(3)}_\mu = - \frac{m^2_\rho}{2\gamma_\rho} \rho_\mu (x) \equiv -
\frac{m^2_\rho}{f_\rho} \rho_\mu (x).
\label{2}
\ee
Consistency of (2) with electromagnetic current conservation requires 
the vector
mesons to be coupled to conserved hadronic currents.

The CFI immediately implies a proportionality between the amplitudes of
interactions induced by real or virtual photons $(\gamma^*)$ and the
corresponding vector-meson-induced processes (e.g. \cite{Schi}),
\be
[\gamma^* A \to B] = -e \frac{m^2_\rho}{2 \gamma_\rho} \frac{1}{q^2-m^2_\rho}
[\rho^0 A \to B] + (\omega) +(\phi).
\label{3}
\ee
According to (3), the photon virtually dissociates, or ``fluctuates'' in
modern jargon, into an on-shell vector meson that subsequently interacts 
with hadron $A$ to yield the hadron state $B$. It is important to note that
the photon fluctuates into an on-shell vector meson, since the dependence 
on the photon virtuality $q^2$ is (by an implicit assumption) solely 
determined by the propagator in (3).

Specializing (3) to $A = B$, we deduce universality of e.g. the 
$\rho^0$-meson coupling, $(f_{\rho AA} = f_{\rho BB} = ... = f_\rho)$
from universality of the electromagnetic coupling,
\be
\frac{e}{2 \gamma_\rho} f_{\rho AA} = \frac{e}{2 \gamma_\rho} f_{\rho BB} =
... = e,
\label{4}
\ee
 connecting the approach of the present section with the one of
section 1, where universality arose as a consequence of the gauge principle.

The CFI, when applied to e.g. the $\gamma^* \to \rho^0 \to 2 \pi$
transition, measurable in $e^+e^-$ annihilation, allows one to express
the coupling constant $\gamma_\rho$ as integral over the $\rho^0$ meson
peak,
\be
\frac{e^2}{4 \gamma^2_\rho} = \frac{\alpha\pi}{\gamma^2_\rho} =
\frac{1}{4 \pi^2 \alpha} \int_{4 m^2_\pi} \sigma_{e^+e^- \to \rho^0 \to
2 \pi} (m^2) dm^2.
\label{5}
\ee
Based on (5), the on-shell vector-meson couplings to the photon were 
accurately measured \cite{N2} as soon as the $e^+ e^-$ storage rings at 
Novosibirsk and Orsay started to produce data around 1966/1967.

The CFI says that the vector mesons are the (only) source of the
Maxwell field, e.g. for the isovector part of the photon interaction,
we have
\be
\partial^\mu F^{(3)}_{\mu \nu} = \frac{e m^2_\rho}{2 \gamma_\rho}
\rho_\nu.
\label{6}
\ee
What is the underlying Lagrangian? A mixing term proportional to
$\rho_\mu A^\mu$ is suggestive, but violates \cite{N3} 
electromagnetic gauge
invariance and yields an imaginary (!) photon mass when summing up
to photon-vector meson transition to all orders in the photon propagator.
The correct form of the Lagrangian, consistent with gauge invariance,
was given by Kroll, Lee, Zumino in 1967 \cite{Kroll}. It may be written in the
``current-mixing'' form (where $J^{(\rho)}_\mu$ is the source of the
$\rho^0$ field),
\be
L_{mix} = - \frac{1}{2} \frac{e}{f_\rho} \rho_{\mu\nu} F^{\mu\nu} +
\frac{e}{f_\rho} A_\mu J^{(\rho) \mu},
\label{7}
\ee
or, equivalently, in the ``mass-mixing'' form,
\be
L^\prime_{mix} = \frac{e^\prime m^2_\rho}{f_\rho} \rho^\prime_\mu 
A^{\prime \mu} - \frac{1}{2} \left( \frac{e^\prime}{f_\rho} \right)^2 m^2_\rho
A^{\prime 2}_\mu
\label{8}
\ee
with $e^2 = e^{\prime 2}/ \left(1 + e^{\prime 2}/f^2_\rho \right)$ and
appropriate linear relations between the fields in (7) and (8).
Recalculating the photon propagator to all orders in the mixing according
to (8) now yields a vanishing photon mass. For most phenomenological
applications of vector-meson dominance, the first term in (8) is
sufficient.

The CFI provides a powerful means to describe photon-hadron interactions.
It is without exaggeration to say that, throughout the 1960ies, it dominated 
our understanding of the electromagnetic interaction of the 
hadrons \cite{Schi, N3a}.

\section{Phenomenological Consequences}
I will concentrate on those phenomenological consequences from vector-meson
dominance that are most relevant for the present-day discussions, in particular
on deep-inelastic scattering (DIS) in the ``diffraction region'' of small
$x \cong Q^2/W^2 << 1$.

\subsection{Vector meson photoproduction, the total photoproduction cross
section, shadowing in photo- and electroproduction from complex nuclei.}

According to the CFI, the amplitudes for photoproduction of vector mesons
and for vector-meson scattering are proportional to each 
other \cite{Schi}. For 
example, for the $\rho^0$ vector meson,
\be
A_{\gamma p \to \rho^0 p} = \frac{e}{2 \gamma_\rho} A_{\rho^0 p \to \rho^0 p}.
\label{9}
\ee
The $\rho^0$-scattering amplitude on the right-hand side in (9) can be
predicted from pion scattering by applying 
\cite{Joos} the additive  quark model,
$\sigma_{\rho^0p} = (1/2) \left(\sigma_{\pi^+ p} + \sigma_{\pi^-p}\right)$.
From (9), vector-meson photoproduction must be weakly dependent on energy
and develop a diffraction peak in the forward direction, as observed in
hadron-hadron interactions. These features, known as ``hadronlike behavior
of the photon'' \cite{Sak,Schi}
were established in the 1960ies by the experiments at
DESY and SLAC.

With vector-meson  dominance applied to the forward Compton scattering 
amplitude, as pointed out by L. Stodolsky \cite{Stodolsky}
in 1967, at sufficiently high energy,
$W$, we have the important sum rule,
\be
\sigma_{\gamma p} (W^2) = \sum_{\rho^0, \omega, \phi} \sqrt{16 \pi}
\sqrt{\frac{\alpha \pi}{\gamma^2_V}} 
\sqrt{\frac{d \sigma^0_{\gamma p \to Vp}}{dt} (W^2)}.
\label{10}
\ee
In (10), for simplicity we have ignored the correction due to the
(small) real part of the vector-meson-production amplitude on
the right-hand side. By the time of the 1971 Conference on Electron
Photon Interactions at Cornell University, it had become clear that
(10) was of approximate validity. The fact that the right-hand side
of (10) yields 78 \% \cite{Wolf}
of the total photoproduction cross section,
$\sigma_{\gamma p} (W^2)$, became the starting
point of Generalized Vector Dominance (GVD) 
\cite{Sa-Schi,Gribov} in 1972\footnote{Compare also ``Extended Vector 
Dominance'' \cite{Bramon}}. The remaining
22 \% were attributed to a continuum of
more massive vector-state contributions, becoming
dominant as soon as the virtuality of the photon becomes large, 
$Q^2 >> m^2_\rho$. I will come back to that.

When hadrons, e.g. pions, are scattered from large complex nuclei,
the nucleons inside the nucleus find themselves in the shadow created
by the ones at the surface, since the mean free path of hadrons in
nuclear matter is smaller than the radius of the nucleus,
$l_h << R$. For very small $l_h$, one expects  hadron-nucleus-interaction 
cross sections to be proportional to the surface of the
nucleus (e.g. \cite{N5})
of mass number $A$, i.e. $\sigma_{h~nucleus} \sim A^{2/3}$.
The actually measured power is slightly larger than $2/3$, since $l_h$
is not sufficiently small compared with $R$.

What is the dependence on $A$ for photon-nucleus interactions? Is it
that $\sigma_{\gamma ~nucleus} \sim A$ or rather $\sigma_{\gamma ~nucleus}
\sim A^{2/3}$, since photons behave hadronlike? The question was posed
and answered by Stodolsky \cite{Stodolsky} in 1967\footnote{Compare also
refs. \cite{N6, N5}.}. 
He realized that the process of
$\gamma$-nucleus scattering must be treated as a two-channel problem,
since the photon in the forward Compton amplitude on nuclei may convert
to a vector meson and reconvert to a photon on a single nucleon as well as on
two different nucleons in the nucleus. The scale that determines the
$A$ dependence may be identified with the lifetime 
\cite{lifetime} of a vector meson
fluctuation. For sufficiently large lifetime, or fluctuation length, $d$,
namely for
\be
d = \frac{2 \nu}{M^2_V} >> R,
\label{11}
\ee
the $A$ dependence becomes hadronlike. Shadowing occurs, provided the
photon energy $\nu$ is sufficiently large. The generalization to virtual
photons reads (e.g. ref. \cite{Bilchak})
\be
d (x, Q^2, m^2_V) = \frac{Q^2}{Mx(Q^2+M^2_V)},
\label{12}
\ee
where $x \cong Q^2/2M\nu$ is the Bjorken scaling variable \cite{N7}.

Shadowing in photoproduction $(Q^2 = 0)$ in the years 1969 to 1973 was 
experimentally found in experiments performed at DESY and SLAC. Compare
Fig. 1 \cite{DS}. After many years of confusion, in 1989, the EMC-NMC 
collaboration
established \cite{EMC}
shadowing in electroproduction $(Q^2 \cong 10 GeV^2)$, compare Fig. 2
\cite{Bilchak}\footnote{See \cite{EMC} also for further references on
the theoretical analysis of the EMC-NMC effect}.
\begin{figure}[htb]
\centerline{\psfig{file=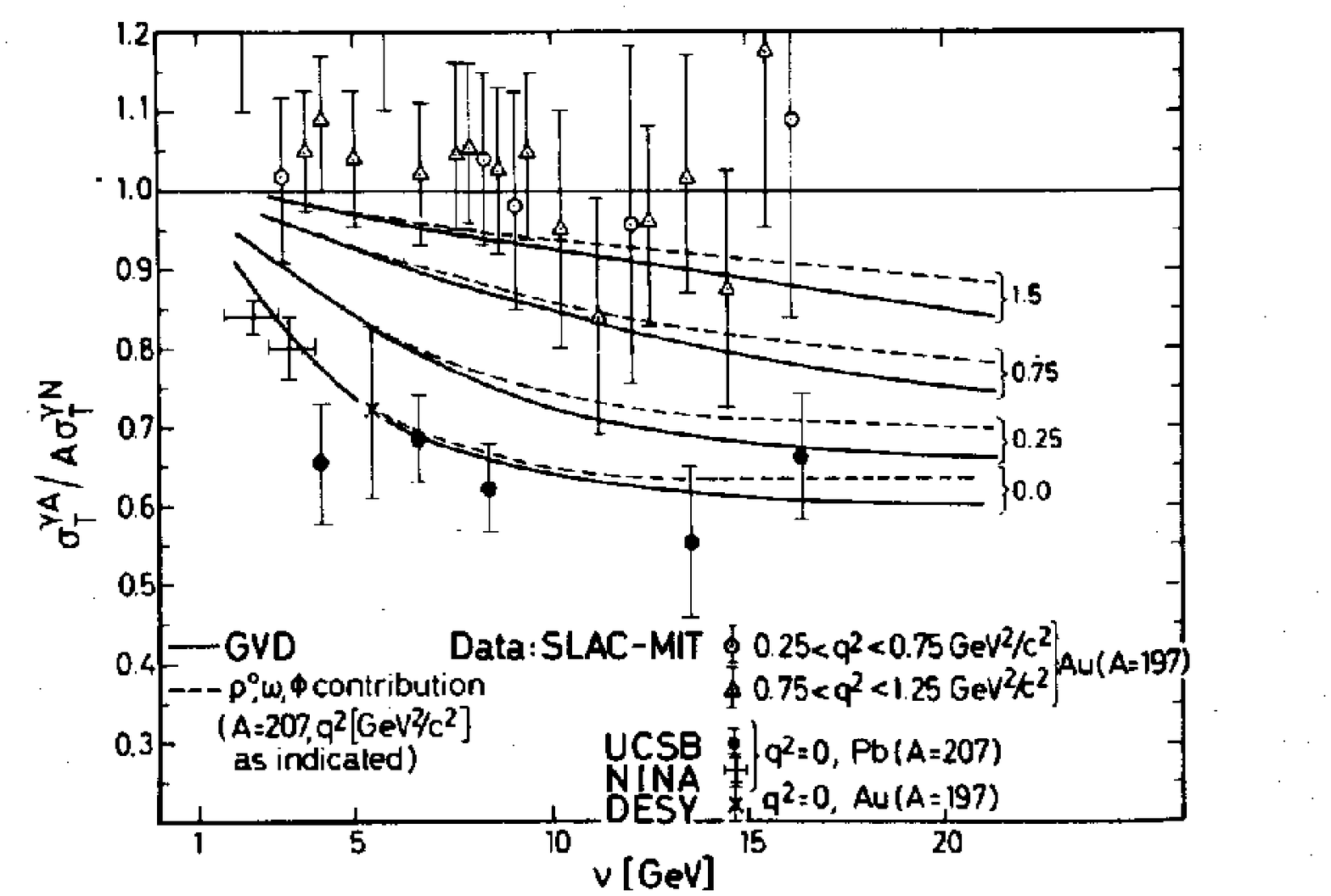, width=6cm}}
\caption{Shadowing in photoproduction (from \cite{DS}).}
\end{figure}

\begin{figure}[htb]
\centerline{\psfig{file=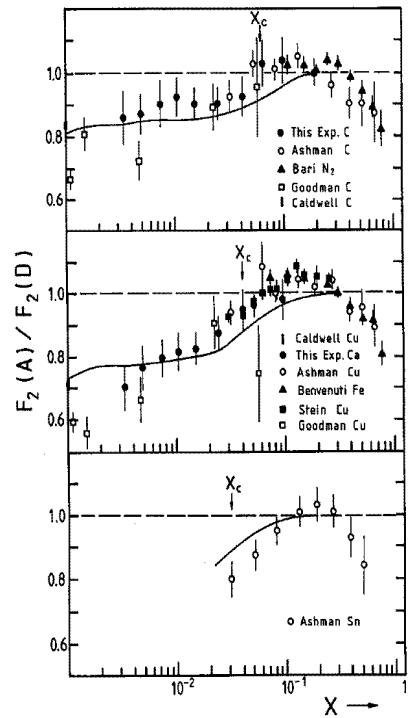, width=6cm,height=10cm}}
\caption{Shadowing in DIS (from \cite{Bilchak}).}
\end{figure}

The EMC-NMC result is of great importance with respect to the present-day
discussions on deep-inelastic scattering (DIS) at low values  of
$x \cong Q^2/W^2$. Since diffractive production and propagation of
$(q \bar q)^{J=1}$ (vector) states is essential for the two-step process
causing shadowing, and the contributions of the low-lying vector mesons
$\rho^0, \omega, \phi$ become negligible at large $Q^2$, the 1989 EMC-NMC
result provided unambiguous evidence for diffractive production of
high-mass $(q \bar q)^{J=1}$ (vector) states in electroproduction prior to
the HERA experiments. Moreover, the shadowing results require those 
high-mass states
to interact hadronlike.

\subsection{$e^+e^-$ annihilation into hadrons, quark-hadron duality.}

In the 1960ies, expectations on how the cross section for $e^+e^- \to
{\rm hadrons}$ would behave at asymptotic energies ranged 
\cite{Sak-Proc} from
$\sigma (e^+ e^- \to {\rm hadrons}) \sim 1/s^3$ as expected from the 
CFI with a finite number of vector mesons, to $\sigma (e^+ e^- \to
{\rm hadrons}) \sim 1/s$. In his 1966 paper \cite{Bjorken} Bjorken said:

{\it ``A speculative argument is presented that the rate of $e^+e^- \to
{\rm hadrons}$ is comparable to the rate of $e^+ e^- \to \mu^+ \mu^-$
in the limit of large energies.''} 

As a consequence of the 1969 SLAC experiment
on DIS, to be summarized below, a scaling behavior of $\sigma (e^+e^- \to
{\rm hadrons}) \sim 1/s$ became stronger acceptance and led to the concept
of quark-hadron duality: the low-lying vector meson peaks are smoothly 
interpolated by $e^+e^-$ annihilation into quark-antiquark pairs,
$e^+e^- \to q \bar q$ of the appropriate flavor \cite{Steiner, Schi-Steiner},
\be
\frac{\alpha \pi}{\gamma^2_V} = \frac{1}{4 \pi^2 \alpha}
\int_{Peak V} ds \sigma_{e^+e^- \to {\rm hadrons}} (s) =
\frac{\alpha R_{e^+e^-}^{(V)}}{3 \pi} \frac{\Delta M^2_V}{M^2_V},
\label{13}
\ee
where $R^{(V)}_{e^+e^-}$ contains the squares of the relevant quark charges.
Compare fig. 3.
Quark-hadron duality became subsequently refined in terms of QCD sum
rules \cite{PR}.

\begin{figure}[htb]
\centerline{\psfig{file=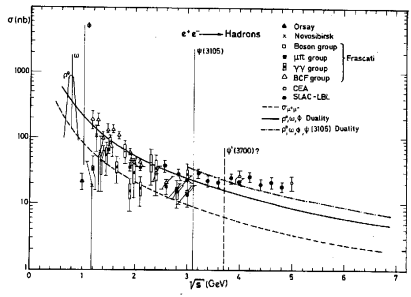, width=6cm}}
\caption{Quark-hadron duality (from \cite{Schi-Steiner}).}
\end{figure}

\subsection{Generalized Vector Dominance and modern picture of DIS
at low $x$.}

The measurements on DIS carried out by the SLAC-MIT collaboration in
1969 \cite{N8} revealed that the transverse part of the photoabsorption cross
section $\sigma_{\gamma^*_T} (W^2, Q^2)$ was decreasing as 
$\sigma_{\gamma^*_T} (W^2, Q^2) \sim 1/Q^2$ in strong disagreement
with the $\rho^0, \omega, \phi$ dominance prediction, where
$\sigma_{\gamma^*_T} (W^2, Q^2) \sim 1/Q^4$. The observed (approximate)
scaling behaviour \cite{N7}
of the structure function $F_2 (x, Q^2) \simeq F_2 (x)$
led Feynman to the parton-model interpretation \cite{Feynman} of the data.

\begin{figure}[htb]
\centerline{\psfig{file=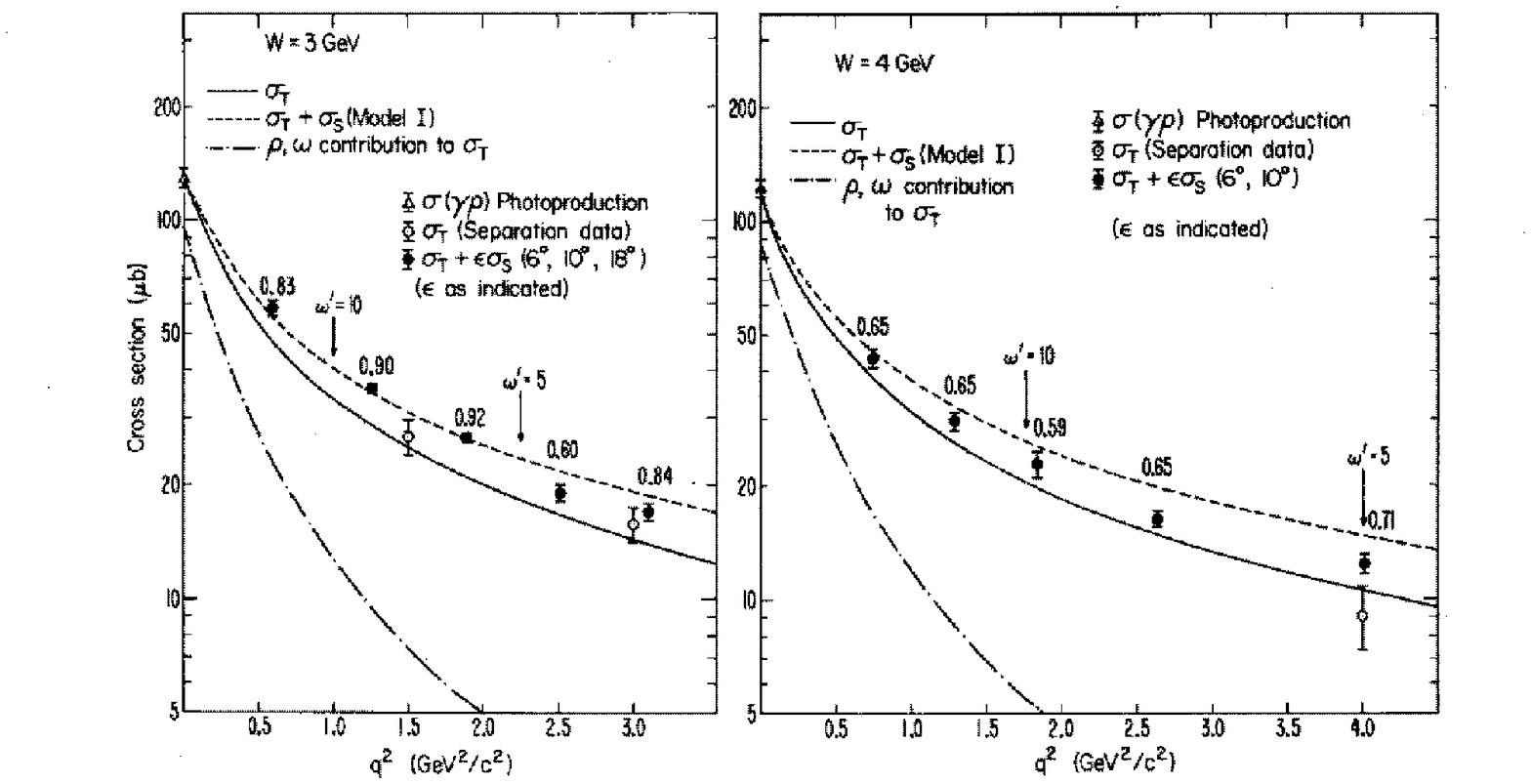, width=8cm}}
\caption{Generalized vector dominance (from \cite{Sa-Schi}).}
\end{figure}

In the generalized vector dominance approach (GVD) \cite{Sa-Schi} 
from 1972, the
coupling of the photon to $\rho^0, \omega, \phi$ mesons was supplemented
by the coupling to a continuum of more massive vector states\footnote{Compare
also the review \cite{N4}. For GVD applied to $\gamma\gamma \to {\rm hadrons}$,
compare \cite{N11}}. A successful
representation of the experimental data in the region of low $x \simeq
Q^2/W^2 \lsim 0.1$ (i.e. large $\omega^\prime$) was obtained, compare fig. 4.
As a verification of the proposed picture, it was concluded \cite{Sa-Schi}
that 

{\it ``Comparisons of higher-mass vector states diffractively produced
in photo- and electroproduction with the final states in $e^+e^-$ collisions
would be of enormous importance.''}

In the SLAC-MIT experiment from 1969 the
region of $x \lsim 0.05$ was not accessible. The exploration of DIS and
diffractive production at low $x$ only started with HERA in 1993.
Consistency requirements between DIS and the suggested scaling in
$e^+ e^-$ annihilation led to the refinement of 
GVD to off-diagonal GVD \cite{Fraas} that 
anticipated a general structure close to the one contained in the modern
approach based on two-gluon exchange from QCD.

The modern picture of DIS at low $x$ \cite{Zakharov} has much in common
\cite{Shoshi,Frankfurt} with (off-diagonal) GVD. The novel element is
the dependence of the virtual forward Compton amplitude on the transverse 
momentum of the exchanged gluon arising from the 
two-gluon-exchange \cite{N9} structure.
The effective value of the transverse momentum of the gluon, $\langle
\vec l^{~2}_\bot \rangle$, introduces a novel scale, characteristic for
DIS at low $x$. The scale is known as ``saturation scale'', $\Lambda^2_{sat}
(W^2)$, since it governs the transition of the total photoabsorption
cross section at fixed $Q^2$ from a strong to a weak energy dependence,
frequently interpreted as an indication for parton 
saturation \cite{N10}. Since the
photon fluctuates to a vector state of two on-shell quarks\footnote{The mass,
the photon fluctuates into, from the analysis of the two-gluon exchange
amplitude is obtained from the photon four momentum, $q^2 = 
\frac{k^2_q + \vec k^{~2}_\bot}{z} + 
\frac{k^2_{\bar q} + \vec k^{~2}_\bot}{1-z}$, by putting $k^2_q = k^2_{\bar q}
= m^2_q$.}, the saturation scale that determines the energy dependence of
the $q \bar q$ color-dipole scattering on the proton depends on the 
energy\footnote{In this respect, among other things, we differ from ref.
\cite{Golec}, where the saturation scale depends on $x$. The energy is
also used as basic variable in ref. \cite{Forshaw}} \cite{Cvetic,Diff2000}.
From the fit with
\be
\Lambda^2_{sat} (W^2) = \frac{1}{6} \langle \vec l^{~2}_\bot \rangle =
B^\prime \left( \frac{W^2}{1 GeV^2} \right)^{C_2}
\ee
we found $C_2^{exp} = 0.27 \pm 0.01,~B^\prime = 0.340 \pm 0.063 GeV^2$.
The experimental data for the total photoabsorption cross section, 
$\sigma_{\gamma^*p} (W^2, Q^2)$ lie on a single curve \cite{Diff2000}
against the novel scaling variable $\eta = (Q^2 + m^2_0)/\Lambda^2_{sat}
(W^2)$. Compare Fig. 5 \cite{Cvetic, Diff2000, MKDS}.
\vspace*{-5cm}
\begin{figure}[htb]
\centerline{\psfig{file=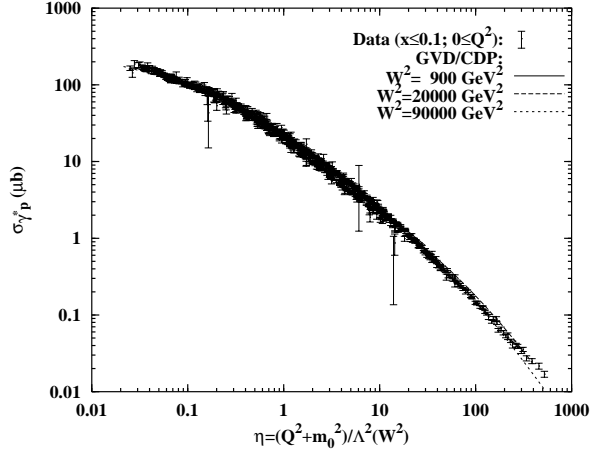, width=8cm}}
\caption{Scaling of $\sigma_{\rho^*p} = \sigma_{\rho^*p} (\eta)$, 
(from \cite{MKDS}).}
\end{figure}

By making use of the duality between the above-mentioned description in
terms of $q \bar q$ scattering and a formulation based on 
$\gamma^*$-quark-gluon scattering in terms of parton distributions, we were
recently able to derive \cite{MKDS}
a theoretical value of $C_2^{theor.} = 0.276$
in surprisingly good agreement with the experimental result. More about
that in the session on DIS and diffraction at this conference.

\section{Conclusions}

In this brief review only a minor part of the relevant photon-hadron
phenomenology could be addressed.

Vector meson physics introduced local gauge transformations to hadron 
interactions. With respect to photons interacting with hadrons, it led to
useful concepts: the fluctuating  photon, hadronlike behavior, 
quark-hadron duality, among others, which stood the test of time.
\vspace*{1cm}

\leftline{\it Acknowledgement}

Many thanks to Maria Krawczyk and her colleagues for the organization
of such a lively and fruitful conference.

This work was supported by Deutsche Forschungsgemeinschaft under
grant No. Schi 189/6-1.

\end{document}